\documentclass{article}
\begin{document}
{\LARGE
\begin{center}
{\bf
${\bf S}$-wave bottom baryons}
\end{center}
}

\large

\begin{center}
\vskip3ex
S.M. Gerasyuta and E.E. Matskevich

\vskip2ex
Department of Theoretical Physics, St. Petersburg State University,
198904,

St. Petersburg, Russia

and

Department of Physics, LTA, 194021, St. Petersburg, Russia
\end{center}

\vskip4ex
\begin{center}
{\bf Abstract}
\end{center}
\vskip4ex
{\large
The masses of $S$-wave bottom baryons are calculated in the framework of
coupled-channel formalism. The relativistic three-quark equations for the
bottom baryons using the dispersion relation technique are found.
The approximate solutions of these equations based on the extraction of
leading singularities of the amplitude are obtained. The calculated mass
values of $S$-wave bottom baryons are in good agreement with the
experimental ones.
\vskip2ex
\noindent
e-mail address: gerasyuta@SG6488.spb.edu

\noindent
e-mail address: matskev@pobox.spbu.ru
\vskip2ex
\noindent
\vskip2ex
{\bf I. Introduction.}
\vskip2ex
Hadron spectroscopy has always played an important role in the revealing
mechanisms underlying the dynamic of strong interactions.

The heavy hadron containing a single heavy quark is particularly
interesting. The light degrees of freedom (quarks and gluons) circle around
the nearby static heavy quark. Such a system behaves as the QCD analogue
of familar hydrogen bound by the electromagnetic interaction.

The heavy quark expansion provides a systematic tool for heavy hadrons.
When the heavy quark mass $m_Q \to \infty$, the angular momentum of the
light degree of freedom is a good quantum number. Therefore heavy hadrons
form doublets. For example, $\Omega_b$ and $\Omega^{*}_b$ will be
degenerate in the heavy quark limit. Their mass splitting is caused by
the chromomagnetic interaction at the order $O(1/m_Q)$, which can be taken
into account systematically in the framework of heavy quark effective
field theory (HQET) [1 -- 3].

Recently CDF Collaboration observed four bottom baryons $\Sigma^{\pm}_b$
and $\Sigma^{* \pm}_b$ [4]. D0 and CDF have seen candidates for $\Xi^-_b$
[5, 6].

In the part two decades, various phenomenological models have been used
to study heavy baryon masses [7 -- 12]. Capstick and Isgur studied the
heavy baryon system in a relativized quark potential model [7]. Roncaglia
et al. predicted the masses of baryons containing one or two heavy quark
using the Feynman-Hellman theorem and semiempirical mass formulas [8].
Jenkins studied heavy baryon masses using a combined expansion of
$1/m_Q$ and $1/N_c$ [9]. Mathur et al. predicted the masses of charmed
and bottom baryons from lattice QCD [10]. Ebert et al. calculated the
masses of heavy baryons with the light-diquark approximation [11].
Stimulated by recent experimental progress, there have been several
theoretical papers on the masses of $\Sigma_b$, $\Sigma^*_b$
and $\Xi_b$ using the hyperfine interaction in the quark model [12 -- 16].
Recently the strong decays of heavy baryons were investigated systematically
using ${ }^3 P_0$ model in Ref. [17].

QCD sum rule has been applied to study heavy baryon masses [18 -- 22].

In our papers [23 -- 25] relativistic generalization of the three-body
Faddeev equations was obtained in the form of dispersion relations in the
pair energy of two interacting particles. The mass spectra of $S$-wave
baryons including $u$, $d$, $s$, $c$ quarks were calculated by a method
based on isolating of the leading singularities in the amplitude.

We searched for the approximate solution of integral three-quark equations
by taking into account two-particle and triangle singularities, the
weaker ones being neglected. If we considered such an approximation, which
corresponds to taking into account two-body and triangle singularities, and
defined all the smooth functions at the middle point of the physical region
of Dalitz-plot, then the problem was reduced to the one of solving a system
of simple algebraic equations.

In the present paper the relativistic three-particle amplitudes in the
coupled-channels formalism are considered. We take into account the
$u$, $d$, $s$, $c$, $b$ quarks and construct the flavor-spin functions
for the $35$ baryons with the spin-parity
$J^p=\frac{1}{2} ^{+}$ and $J^p=\frac{3}{2} ^{+}$:

\vskip2ex

$$
\begin{tabular}{llp{5cm}ll}
 & $J^p=\frac{1}{2} ^{+}$ & & & $J^p=\frac{3}{2} ^{+}$ \\
 & & & & \\
$\Sigma_b$ & $uub, udb, ddb$ & & $\Sigma_b$ & $uub, udb, ddb$ \\
$\Lambda_b$ & $udb$ & & $\Xi_{sb}$ & $usb, dsb$ \\
$\Xi_{sb}^A$ & $usb, dsb$ & & $\Omega_{ssb}$ & $ssb$ \\
$\Xi_{sb}^S$ & $usb, dsb$ & & $\Xi_{cb}$ & $ucb, dcb$ \\
$\Omega_{ssb}$ & $ssb$ & & $\Omega_{scb}$ & $scb$ \\
$\Xi_{cb}^A$ & $ucb, dcb$ & & $\Omega_{ccb}$ & $ccb$ \\
$\Xi_{cb}^S$ & $ucb, dcb$ & & $\Xi_{bb}$ & $ubb, dbb$ \\
$\Lambda_{scb}^A$ & $scb$ & & $\Omega_{sbb}$ & $sbb$ \\
$\Lambda_{scb}^S$ & $scb$ & & $\Omega_{cbb}$ & $cbb$ \\
$\Omega_{ccb}$ & $ccb$ & & $\Omega_{bbb}$ & $bbb$ \\
$\Xi_{bb}$ & $ubb, dbb$ & & & \\
$\Omega_{sbb}$ & $sbb$ & & & \\
$\Omega_{cbb}$ & $cbb$ & & & \\
\end{tabular}
\eqno (1)$$

\vskip2ex

In the paper [25] the relativistic equations were obtained and the mass
spectrum of $S$-wave charmed baryons was calculated.

In the present paper we will be able to use the similar method. In this
case we consider $35$ baryons with the spin-parity
$J^p=\frac{1}{2} ^{+}$ and $J^p=\frac{3}{2} ^{+}$, which include one, two
and three bottom quarks. We have considered the $23$ baryons with different
masses.

The paper is organized as follows. In Section II we obtain the relativistic
three-particle equations which describe the interaction of the quarks in
baryons. In Section III the coupled systems of equations for the reduced
amplitudes are derived. Section IV is devoted to a discussion of the results
for the mass spectrum of $S$-wave bottom baryons (Tables I, II). In the
conclusion the status of the considered model is discussed.

In Appendix A we write down the three-particle integral equations for the
$J^p=\frac{1}{2} ^{+}$ and $J^p=\frac{3}{2} ^{+}$ bottom baryon multiplets.
In Appendix B the coupled systems of approximate equations for lowest
bottom baryons are given.

\newpage
{\bf II. The three-quark integral equations for the ${\bf S}$-wave bottom
baryons.}
\vskip2ex

We calculate the masses of the bottom baryons in a relativistic approach
using the dispersion relation technique. The relativistic three-quark
integral equations are constructed in the form of the dispersion relations
over the two-body subenergy.

We use the graphic equations for the functions $A_J(s,s_{ik})$ [23 -- 25].
In order to represent the amplitude $A_J(s,s_{ik})$ in the form
of dispersion relations, it is necessary to define the amplitudes of
quark-quark interaction $a_J(s_{ik})$. The pair amplitudes
$qq\rightarrow qq$ are calculated in the framework of the dispersion
$N/D$ method with the input four-fermion interaction with quantum numbers
of the gluon [26]. We use results of our relativistic quark model [27]
and write down the pair quark amplitudes in the following form

$$a_J(s_{ik})=\frac{G^2_J(s_{ik})}
{1-B_J(s_{ik})},\eqno (2)$$

$$B_J(s_{ik})=\int\limits_{(m_i+m_k)^2}^{\Lambda_J (i,k)}\hskip2mm
\frac{ds'_{ik}}{\pi}\frac{\rho_J(s'_{ik})G^2_J(s'_{ik})}
{s'_{ik}-s_{ik}},\eqno (3)$$

$$\rho_J (s_{ik})=\frac{(m_i+m_k)^2}{4\pi}
\left(\alpha_J\frac{s_{ik}}{(m_i+m_k)^2}
+\beta_J +\frac{\delta_J}{s_{ik}} \right)\times$$

$$\times\frac{\sqrt{(s_{ik}-(m_i+m_k)^2)(s_{ik}-(m_i-m_k)^2)}}
{s_{ik}}\, .\eqno (4)$$

Here $G_J$ is the vertex function of a diquark, which can be expressed in
terms of the $N$-function of the bootstrap $N/D$ method as $G_J=\sqrt{N_J}$;
$B_J(s_{ik})$ is the Chew-Mandelstam function [28], and $\rho_J (s_{ik})$
is the phase space for a diquark. $s_{ik}$ is the pair energy squared of
diquark, the index $J^p$ determines the spin-parity of diquark.
The coefficients of Chew-Mandelstam function $\alpha_J$, $\beta_J$ and
$\delta_J$ in Table III are given. $\Lambda_J(i,k)$ is the pair energy
cutoff. In the case under discussion the interacting pairs of quarks do not
form bound states. Therefore the integration in the dispersion integral (3)
is carried out from $(m_i+m_k)^2$ to $\Lambda_J(i,k)$ (i,k=1,2,3). Including
all possible rescatterings of each pair of quarks and grouping the terms
according to the final states of the particles, we obtained the coupled
systems of integral equations. For instance, for the
$\Sigma^+_b$ with $J^p=\frac{1}{2} ^{+}$ the wave function is
$\varphi_{\Sigma^+_b}=\sqrt{\frac{2}{3}}\{u\uparrow d\uparrow b\downarrow\}-
\sqrt{\frac{1}{6}}\{u\uparrow d\downarrow b\uparrow\}-
\sqrt{\frac{1}{6}}\{u\downarrow d\uparrow b\uparrow\}$. Then the coupled
system of equations has the following form:

$$\left\{
\begin{array}{l}
A_1(s,s_{12})=\lambda\, b_1(s_{12})L_1(s_{12})+
K_1(s_{12})\left[\frac{1}{4}A_{1^b}(s,s_{13})+
\frac{3}{4}A_{0^b}(s,s_{13})+\right.\\
\\
\hskip10ex \left.
+\frac{1}{4}A_{1^b}(s,s_{23})+\frac{3}{4}A_{0^b}(s,s_{23})
\right]\\
\\
A_{1^b}(s,s_{13})=\lambda\, b_{1^b}(s_{13})L_{1^b}(s_{13})+
K_{1^b}(s_{13})\left[\frac{1}{2}A_1(s,s_{12})-
\frac{1}{4}A_{1^b}(s,s_{12})+\right.\\
\\
\hskip10ex \left.
+\frac{3}{4}A_{0^b}(s,s_{12})+\frac{1}{2}A_1(s,s_{23})-
\frac{1}{4}A_{1^b}(s,s_{23})+\frac{3}{4}A_{0^b}(s,s_{23})
\right] \\
\\
A_{0^b}(s,s_{23})=\lambda\, b_{0^b}(s_{23})L_{0^b}(s_{23})+
K_{0^b}(s_{23})\left[\frac{1}{2}A_1(s,s_{12})+
\frac{1}{4}A_{1^b}(s,s_{12})+\right.\\
\\
\hskip10ex \left.
+\frac{1}{4}A_{0^b}(s,s_{12})+\frac{1}{2}A_1(s,s_{13})+
\frac{1}{4}A_{1^b}(s,s_{13})+\frac{1}{4}A_{0^b}(s,s_{13})
\right] \, .\\
\end{array} \right.
\eqno (5)$$

\noindent
Here the function $L_J(s_{ik})$ has the form

$$L_J(s_{ik})=\frac{G_J(s_{ik})}{1-B_J(s_{ik})}.\eqno (6)$$

\noindent
The integral operator $K_J (s_{ik})$ is

$$K_J (s_{ik})=L_J(s_{ik})\, \int\limits_{(m_i+m_k)^2}^{\Lambda_J(ik)}
\hskip2mm\frac{ds'_{ik}}{\pi}\frac{\rho_J(s'_{ik})G_J(s'_{ik})}
{s'_{ik}-s_{ik}}\, \int\limits_{-1}^{1}\frac{dz}{2}\, .\eqno (7)$$

\noindent
The function $b_J(s_{ik})$ is the truncated function of Chew-Mandelstam:

$$b_J(s_{ik})=\int\limits_{(m_i+m_k)^2}^{\infty}\hskip2mm
\frac{ds'_{ik}}{\pi}\frac{\rho_J(s'_{ik})G_J(s'_{ik})}
{s'_{ik}-s_{ik}},\eqno (8)$$

\noindent
$z$ is the cosine of the angle between the relative momentum of particles
$i$ and $k$ in the intermediate state and the momentum of
particle $j$ in the final state, taken in the c.m. of the particles
$i$ and $k$. Let some current produces three quarks
with the vertex constant $\lambda$. This constant do not affect to the
spectra mass of bottom baryons.
By analogy with the $\Sigma^+_b$ state we obtain
the rescattering amplitudes of the three various quarks for the all
bottom states (Appendix A).

\vskip2ex
{\bf III. Reduced equations for the ${\bf S}$-wave bottom baryons.}
\vskip2ex

Let us extract two-particle singularities in $A_J(s,s_{ik})$:

$$A_J(s,s_{ik})=\frac{\alpha_J(s,s_{ik})b_J(s_{ik})G_J(s_{ik})}
{1-B_J(s_{ik})},\eqno (9)$$

\noindent
$\alpha_J(s,s_{ik})$ is the reduced amplitude. Accordingly to this,
all integral equations can be rewritten using the reduced amplitudes.
The function $\alpha_J(s,s_{ik})$ is the smooth function of $s_{ik}$
as compared with the singular part of the amplitude. We do not extract
the three-body singularities, because they are weaker than the
two-particle singularities. For instance, one considers the first equation
of system for the $\Sigma^+_b$ with $J^p=\frac{1}{2}^+$:

$$\alpha_1 (s,s_{12})=\lambda+\frac{1}{b_1(s_{12})}
\, \int\limits_{(m_1+m_2)^2}^{\Lambda_1(1,2)}\hskip2mm
\frac{ds'_{12}}{\pi}\,\frac{\rho_1(s'_{12})G_1(s'_{12})}
{s'_{12}-s_{12}}\times$$

$$\times\int\limits_{-1}^{1}\frac{dz}{2}\,
\left(
\frac{G_{1^b}(s'_{13})b_{1^b}(s'_{13})}{1-B_{1^b}(s'_{13})}
\,\frac{1}{2}\,\alpha_{1^b}(s,s'_{13})+
\frac{G_{0^b}(s'_{13})b_{0^b}(s'_{13})}{1-B_{0^b}(s'_{13})}
\,\frac{3}{2}\,\alpha_{0^b}(s,s'_{13})
\right).\eqno (10)$$

The connection between $s'_{12}$ and $s'_{13}$ is [29]:

$$s'_{13}=m_1^2+m_3^2-\frac{\left(s'_{12}+m_3^2-s\right)
\left(s'_{12}+m_1^2-m_2^2\right)}{2s'_{12}}\pm$$

$$\pm\frac{z}{2s'_{12}}\times\sqrt{\left(s'_{12}-(m_1+m_2)^2\right)
\left(s'_{12}-(m_1-m_2)^2\right)}$$

$$\times\sqrt{\left(s'_{12}-(\sqrt{s}+m_3)^2\right)
\left(s'_{12}-(\sqrt{s}-m_3)^2\right)}\, .\eqno (11)$$

The formula for $s'_{23}$ is similar to (11) with replaced by $z \to -z$.
Thus $A_{1^b}(s,s'_{13})+A_{1^b}(s,s'_{23})$ must be replaced by
$2A_{1^b}(s,s'_{13})$. $\Lambda_J(i,k)$ is the cutoff at the large
value of $s_{ik}$, which determines the contribution from small distances.

The construction of the approximate solution of coupled system equations
is based on the extraction of the leading singularities which are close to
the region $s_{ik}=(m_i+m_k)^2$ [29].

We consider the approximation, which corresponds to the single interaction
of the all three particles (two-particle and triangle singularities) and
neglecting all the weaker ones.

The functions $\alpha_J(s,s_{ik})$ are the smooth functions of $s_{ik}$
as compared with the singular part of the amplitude, hence it can be
expanded in a series at the singulary point and only the first term of
this series should be employed further. As $s_0$ it is convenient to
take the middle point of physical region of the Dalitz plot in which $z=0$.
In this case we get from (11)
$s_{ik}=s_0=\frac{s+m_1^2+m_2^2+m_3^2}{m_{12}^2+m_{13}^2+m_{23}^2}$,
where $m_{ik}=\frac{m_i+m_k}{2}$. We define functions $\alpha_J(s,s_{ik})$
and $b_J(s_{ik})$ at the point $s_0$. Such a choice of point $s_0$ allows us
to replace integral equations (5) by the algebraic couple equations for the
state $\Sigma^+_b$:

$$\left\{
\begin{array}{l}
\alpha_1(s,s_0)=\lambda+\frac{1}{2}\,\alpha_{1^b}(s,s_0)
\, I_{1 1^b}(s,s_0)\,\frac{b_{1^b}(s_0)}{b_1(s_0)}
+\frac{3}{2}\,\alpha_{0^b}(s,s_0)\, I_{1 0^b}(s,s_0)
\,\frac{b_{0^b}(s_0)}{b_1(s_0)}\\
\\
\alpha_{1^b}(s,s_0)=\lambda
+\alpha_1(s,s_0)\, I_{1^b 1}(s,s_0)\,\frac{b_1(s_0)}{b_{1^b}(s_0)}\\
\\
\hskip10ex
-\frac{1}{2}\,\alpha_{1^b}(s,s_0)\, I_{1^b 1^b}(s,s_0)
+\frac{3}{2}\,\alpha_{0^b}(s,s_0)\, I_{1^b 0^b}(s,s_0)
\,\frac{b_{0^b}(s_0)}{b_{1^b}(s_0)}\\
\\
\alpha_{0^b}(s,s_0)=\lambda+\alpha_1(s,s_0)\, I_{0^b 1}(s,s_0)
\,\frac{b_1(s_0)}{b_{0^b}(s_0)}\\
\\
\hskip10ex
+\frac{1}{2}\,\alpha_{1^b}(s,s_0)\, I_{0^b 1^b}(s,s_0)
\,\frac{b_{1^b}(s_0)}{b_{0^b}(s_0)}
+\frac{1}{2}\,\alpha_{0^b}(s,s_0)\, I_{0^b 1^b}(s,s_0)
\, .\\
\end{array} \right.\eqno (12)$$

The function $I_{J_1 J_2}(s,s_0)$ takes into account singularity
which corresponds to the simultaneous vanishing of all propagators in the
triangle diagram.

$$I_{J_1 J_2}(s,s_0)=\int\limits_{(m_i+m_k)^2}^{\Lambda_{J_1}}\hskip2mm
\frac{ds'_{ik}}{\pi}\frac{\rho_{J_1}(s'_{ik})G^2_{J_1}(s'_{ik})}
{s'_{ik}-s_{ik}}\, \int\limits_{-1}^{1}\frac{dz}{2}\,
\frac{1}{1-B_{J_2}(s_{ij})}\eqno (13)$$

The $G_J(s_{ik})$ functions have the smooth dependence from energy
$s_{ik}$ [27] therefore we suggest them as constants. The parameters of
model: $g_J$ vertex constant and $\lambda_J$ cutoff parameter are chosen
dimensionless.

$$g_J=\frac{m_i+m_k}{2\pi}G_J , \,\,\, \lambda_J=\frac{4\Lambda_J}
{(m_i+m_k)^2}.\eqno (14)$$

Here $m_i$ and $m_k$ are quark masses in the intermediate state of the quark
loop. We calculate the coupled system of equations and can determine the
mass values of the $\Sigma^+_b$ state. We calculate a pole in $s$
which corresponds to the bound state of the three quarks.

By analogy with $\Sigma^+_b$-hyperon we obtain the systems of equations for
the reduced amplitudes of all bottom baryons (Appendix B).

The solutions of the coupled system of equations are considered as:

$$\alpha_J=\frac{F_J(s,\lambda_J)}{D(s)}\, , \eqno(15)$$

\noindent
where the zeros of the $D(s)$ determinate the masses of bound
states of baryons. $F_J(s,\lambda_J)$ are the functions of $s$ and
$\lambda_J$. The functions $F_J(s,\lambda_J)$ determine the contributions
of subamplitudes to the baryon amplitude.

\vskip2ex
{\bf IV. Calculation results.}
\vskip2ex

The quark masses ($m_u=m_d=m$, $m_s$ and $m_c$) are given similar to the
our paper ones [25]: $m=0.495\, GeV$, $m_s=0.77\, GeV$ and
$m_c=1.655\, GeV$. The strange quark mass is heavier than the strange quark
mass in the some quark models [7 -- 11, 23, 27]. This value of strange quark
mass allows us to describe the spectroscopy of $S$-wave charmed baryons
well [25]. We use only two new parameters as compared with the $S$-wave
light baryons [23, 24]. The parameters of model are the cutoff energy
parameters $\lambda_q=10.7$, $\lambda_c=6.5$ for the light $u$, $d$, $s$
and charmed quarks, the vertex constants $g_0=0.70$, $g_1=0.55$, for the
light diquarks with $J^P=0^+$, $1^+$ and $g_c=0.857$ for the charmed
diquarks. $\lambda_{qQ}=\frac{1}{4}(\sqrt{\lambda_q}+\sqrt{\lambda_Q})^2$
are chosen ($q=u, d, s$, $Q=c$).

In the present paper we have used two new parameters: the cutoff of the $bb$
diquark  $\lambda_b=5.4$ and the coupling constant $g_b=1.03$.
These values have been determined by the $b$-baryon masses:
$M_{\Sigma_b \frac{1}{2}^+}=5.808\, GeV$ and
$M_{\Sigma_b \frac{3}{2}^+}=5.829\, GeV$. In order to fix $m_b=4.840\, GeV$
we use the $b$-baryon masses $M_{\Sigma_b \frac{3}{2}^+}=5.829\, GeV$.
We represent the masses of all $S$-wave bottom baryons in the Tables I, II.
The calculated mass value $M_{\Lambda_b \frac{1}{2}^+}=5.624\, GeV$ is equal
to the experimental data [30], the mass value
$M_{\Xi^A_{sb} \frac{1}{2}^+}=5.761\, GeV$ is close to the experimental
one [5]. But the more precise CDF mass (Table I) lies close to a prediction
of Ref. [13]. We neglect with the mass distinction of $u$ and $d$ quarks.
The estimation of the theoretical error on the bottom baryon masses is
$2-5\, MeV$. This result was obtained by the choice of model parameters.

\vskip2ex
{\bf V. Conclusion.}
\vskip2ex

In a strongly bound systems, which include the light quarks, where
$p/m \sim 1$, the approximation by nonrelativistic kinematics and dynamics
is not justified.

In our paper the relativistic description of three particles amplitudes
of bottom baryons are considered. We take into account the $u$, $d$, $s$,
$c$, $b$ quarks. The mass spectrum of $S$-wave bottom baryons with one,
two and three $b$ quarks is considered. We use only two new parameters
for the calculation of $23$ baryon masses. The other model parameters
in the our papers [23 -- 25] are given. The charge-averaged hyperfine
splitting between the $J=\frac{1}{2}$ and $J=\frac{3}{2}$ states predicted
from that for charmed particles is similar to the Ref. [31].

In their paper the spin-averaged mass of the states $\Xi'_b$ and $\Xi^*_b$
is predicted to lie around to $150-160\, MeV$ above $M_{\Xi_b}$, while the
hyperfine splitting between $\Xi'_b$ and $\Xi^*_b$ is predicted to lie in
the rough range of $20$ to $30\, MeV$.

We have predicted the masses of baryons containing $b$ quarks using the
coupled-channel formalism. We believe that the prediction for the $S$-wave
bottom baryons based on the relativistic kinematics and dynamics allows as
to take into account the relativistic corrections. In our consideration the
bottom baryon masses are heavier than the masses in the other quark models
[11, 31 -- 34]. In our model the spin-averaged mass of the states $\Xi'_b$
and $\Xi^*_b$ is predicted to lie around to $250\, MeV$ above $M_{\Xi_b}$.
The relativistic corrections are particularly important for the splitting
between $\Omega^+_b$ and $\Omega_b$ baryons.

We will be able to calculate the $P$-wave bottom baryons in our approach
[35, 36] using the new experimental data. The interesting opinions with
the $S$-matrix singularities in Ref. [37] are given.

\vskip2.0ex
{\bf Acknowledgments.}
\vskip2.0ex

The authors would like to thank S. Capstick and S.L. Zhu for useful
discussions. The work was carried with the support
of the Russion Ministry of Education (grant 2.1.1.68.26).

\newpage
{\bf Appendix A. Integral equations for three-particle amplitudes
${\bf A_{J^p}(s, s_{ik})}$ of ${\bf S}$-wave bottom baryons.}
\vskip2ex
{\bf The $J^P=\frac{3}{2}^+$ multiplet.}
\vskip2ex

\noindent
1. Baryons $\Sigma_b$, $\Omega_{ssb}$, $\Omega_{ccb}$, $\Xi_{bb}$,
$\Omega_{sbb}$, $\Omega_{cbb}$.

\noindent
The wave functions: $\varphi=\{x\uparrow x\uparrow y\uparrow\}$.

$$\left\{
\begin{array}{l}
A_{1^{xx}}(s,s_{12})=\lambda\, b_{1^{xx}}(s_{12})L_{1^{xx}}(s_{12})+
K_{1^{xx}}(s_{12})\left[A_{1^{xy}}(s,s_{13})+A_{1^{xy}}(s,s_{23})\right]\\
\\
A_{1^{xy}}(s,s_{13})=\lambda\, b_{1^{xy}}(s_{13})L_{1^{xy}}(s_{13})+
K_{1^{xy}}(s_{13})\left[A_{1^{xx}}(s,s_{12})+A_{1^{xy}}(s,s_{23})\right]
\, .\\
\end{array}
\right.\eqno (A1)$$

\noindent
Here: for the $\Sigma_b$: $x=q$ $(q=u, d)$, $y=b$;
for the $\Omega_{ssb}$: $x=s$, $y=b$; for the $\Omega_{ccb}$: $x=c$, $y=b$;
for the $\Xi_{bb}$: $x=b$, $y=q$; for the $\Omega_{sbb}$: $x=b$, $y=s$;
for the $\Omega_{sbb}$: $x=b$, $y=c$.

\vskip2ex

\noindent
2. Baryons $\Xi_{sb}$, $\Xi_{cb}$, $\Omega_{scb}$.

\noindent
The wave functions: $\varphi=\{x\uparrow y\uparrow z\uparrow\}$.

$$\left\{
\begin{array}{l}
A_{1^{xy}}(s,s_{12})=\lambda\, b_{1^{xy}}(s_{12})L_{1^{xy}}(s_{12})+
K_{1^{xy}}(s_{12})\left[A_{1^{xz}}(s,s_{13})+A_{1^{yz}}(s,s_{23})\right]\\
\\
A_{1^{xz}}(s,s_{13})=\lambda\, b_{1^{xz}}(s_{13})L_{1^{xz}}(s_{13})+
K_{1^{xz}}(s_{13})\left[A_{1^{xy}}(s,s_{12})+A_{1^{yz}}(s,s_{23})\right]\\
\\
A_{1^{yz}}(s,s_{23})=\lambda\, b_{1^{yz}}(s_{23})L_{1^{yz}}(s_{23})+
K_{1^{yz}}(s_{23})\left[A_{1^{xy}}(s,s_{12})+A_{1^{xz}}(s,s_{13})\right]
\, .\\
\end{array}
\right.\eqno (A2)$$

\noindent
Here: for the $\Xi_{sb}$: $x=q$, $y=s$, $z=b$;
for the $\Xi_{cb}$: $x=q$, $y=c$, $z=b$;
for the $\Omega_{scb}$: $x=s$, $y=c$, $z=b$.

\vskip2ex

\noindent
3. Baryon $\Omega_{bbb}$.

\noindent
The wave functions: $\varphi=\{b\uparrow b\uparrow b\uparrow\}$.

$$
\begin{array}{l}
A_{1^{bb}}(s,s_{12})=\lambda\, b_{1^{bb}}(s_{12})L_{1^{bb}}(s_{12})+
K_{1^{bb}}(s_{12})\left[A_{1^{bb}}(s,s_{13})+A_{1^{bb}}(s,s_{23})\right]
\, .\\
\end{array}
\eqno (A3)$$


\vskip2ex
{\bf The $J^P=\frac{1}{2}^+$ multiplet.}
\vskip2ex

\noindent
1. Baryons $\Sigma_b$, $\Lambda_b$, $\Omega_{ssb}$, $\Omega_{ccb}$,
$\Xi_{bb}$, $\Omega_{sbb}$, $\Omega_{cbb}$.

\noindent
The wave functions:

\noindent
for the $\Sigma_b$:

$\varphi_{\Sigma^+_b}=\sqrt{\frac{2}{3}}\{u\uparrow d\uparrow b\downarrow\}-
\sqrt{\frac{1}{6}}\{u\uparrow d\downarrow b\uparrow\}-
\sqrt{\frac{1}{6}}\{u\downarrow d\uparrow b\uparrow\}$;

\noindent
for the $\Lambda_b$:

$\varphi_{\Lambda_b}=\sqrt{\frac{1}{2}}\{u\uparrow d\downarrow b\uparrow\}-
\sqrt{\frac{1}{2}}\{u\downarrow d\uparrow b\uparrow\}$;

\noindent
for the $\Omega_{ssb}$, $\Omega_{ccb}$, $\Xi_{bb}$, $\Omega_{sbb}$,
$\Omega_{cbb}$:

$\varphi=\sqrt{\frac{2}{3}}\{x\uparrow x\uparrow y\downarrow\}-
\sqrt{\frac{2}{3}}\{x\uparrow x\downarrow y\uparrow\}$;

\noindent
here: for the $\Omega_{ssb}$: $x=s$, $y=b$; for the $\Omega_{ccb}$: $x=c$,
$y=b$; for the $\Xi_{bb}$: $x=b$, $y=q$; for the $\Omega_{sbb}$: $x=b$,
$y=s$; for the $\Omega_{cbb}$: $x=b$, $y=c$.

$$\left\{
\begin{array}{l}
A_x(s,s_{12})=\lambda\, b_x(s_{12})L_x(s_{12})+
K_x(s_{12})\left[\frac{1}{4}A_y(s,s_{13})+
\frac{3}{4}A_z(s,s_{13})+\right.\\
\\
\hskip10ex \left.
+\frac{1}{4}A_y(s,s_{23})+\frac{3}{4}A_z(s,s_{23})
\right]\\
\\
A_y(s,s_{13})=\lambda\, b_y(s_{13})L_y(s_{13})+
K_y(s_{13})\left[\frac{1}{2}A_x(s,s_{12})-
\frac{1}{4}A_y(s,s_{12})+\right.\\
\\
\hskip10ex \left.
+\frac{3}{4}A_z(s,s_{12})+\frac{1}{2}A_x(s,s_{23})-
\frac{1}{4}A_y(s,s_{23})+\frac{3}{4}A_z(s,s_{23})
\right] \\
\\
A_z(s,s_{23})=\lambda\, b_z(s_{23})L_z(s_{23})+
K_z(s_{23})\left[\frac{1}{2}A_x(s,s_{12})+
\frac{1}{4}A_y(s,s_{12})+\right.\\
\\
\hskip10ex \left.
+\frac{1}{4}A_z(s,s_{12})+\frac{1}{2}A_x(s,s_{13})+
\frac{1}{4}A_y(s,s_{13})+\frac{1}{4}A_z(s,s_{13})
\right] \, .\\
\end{array}
\right.\eqno (A4)$$

\noindent
Here: for the $\Sigma_b$: $x=1^{qq}$, $y=1^{qb}$, $z=0^{qb}$;
for the $\Lambda_b$: $x=0^{qq}$, $y=0^{qb}$, $z=1^{qb}$;
for the $\Omega_{ssb}$: $x=1^{ss}$, $y=1^{sb}$, $z=0^{sb}$;
for the $\Omega_{ccb}$: $x=1^{cc}$, $y=1^{cb}$, $z=0^{cb}$;
for the $\Xi_{bb}$: $x=1^{bb}$, $y=1^{qb}$, $z=0^{qb}$;
for the $\Omega_{sbb}$: $x=1^{bb}$, $y=1^{sb}$, $z=0^{sb}$;
for the $\Omega_{cbb}$: $x=1^{bb}$, $y=1^{cb}$, $z=0^{cb}$.

\vskip2ex

\noindent
2. Baryons $\Xi_{sb}^A$, $\Xi_{sb}^S$, $\Lambda_{scb}^A$, $\Lambda_{scb}^S$,
$\Xi_{cb}^A$, $\Xi_{cb}^S$.

\noindent
The wave functions:

\noindent
for the $\Xi_{sb}^A$, $\Lambda_{scb}^A$, $\Xi_{cb}^A$:

$\varphi=\sqrt{\frac{1}{2}}\{x\uparrow y\uparrow z\downarrow\}-
\sqrt{\frac{1}{2}}\{x\uparrow y\downarrow z\uparrow\}$;

\noindent
here: for the $\Xi_{sb}^A$: $x=b$, $y=s$, $z=q$;
for the $\Lambda_{scb}^A$: $x=b$, $y=c$, $z=s$;
for the $\Xi_{cb}^A$: $x=b$, $y=c$, $z=q$.

\noindent
For the $\Xi_{sb}^S$, $\Lambda_{scb}^S$, $\Xi_{cb}^S$:

$\varphi=\sqrt{\frac{2}{3}}\{x\uparrow y\uparrow z\downarrow\}-
\sqrt{\frac{1}{6}}\{x\uparrow y\downarrow z\uparrow\}-
\sqrt{\frac{1}{6}}\{x\downarrow y\uparrow z\uparrow\}$;

\noindent
here: for the $\Xi_{sb}^S$: $x=q$, $y=s$, $z=b$;
for the $\Lambda_{scb}^S$: $x=s$, $y=c$, $z=b$;
for the $\Xi_{cb}^S$: $x=q$, $y=c$, $z=b$.

$$\left\{
\begin{array}{l}
A_x(s,s_{12})=\lambda\, b_x(s_{12})L_x(s_{12})+
K_x(s_{12})\left[\frac{1}{8}A_y(s,s_{13})+
\frac{1}{8}A_z(s,s_{13})+
\right.\\
\\
\hskip10ex
+\frac{3}{8}A_v(s,s_{13})+\frac{3}{8}A_w(s,s_{13})
+\frac{1}{8}A_y(s,s_{23})+\frac{1}{8}A_z(s,s_{23})+\\
\\
\hskip10ex \left.
+\frac{3}{8}A_v(s,s_{23})+\frac{3}{8}A_w(s,s_{23})
\right]\\
\\
A_y(s,s_{13})=\lambda\, b_y(s_{13})L_y(s_{13})+
K_y(s_{13})\left[\frac{1}{2}A_x(s,s_{12})
-\frac{1}{4}A_z(s,s_{12})+
\right.\\
\\
\hskip10ex \left.
+\frac{3}{4}A_w(s,s_{12})+\frac{1}{2}A_x(s,s_{23})
-\frac{1}{4}A_z(s,s_{23})+\frac{3}{4}A_w(s,s_{23})\right]\\
\\
A_z(s,s_{23})=\lambda\, b_z(s_{23})L_z(s_{23})+
K_z(s_{23})\left[\frac{1}{2}A_x(s,s_{12})
-\frac{1}{4}A_y(s,s_{12})+
\right.\\
\\
\hskip10ex \left.
+\frac{3}{4}A_v(s,s_{12})+\frac{1}{2}A_x(s,s_{13})
-\frac{1}{4}A_y(s,s_{13})+\frac{3}{4}A_v(s,s_{13})\right]\\
\\
A_v(s,s_{13})=\lambda\, b_v(s_{13})L_v(s_{13})+
K_v(s_{13})\left[\frac{1}{2}A_x(s,s_{12})
+\frac{1}{4}A_z(s,s_{12})+
\right.\\
\\
\hskip10ex \left.
+\frac{1}{4}A_w(s,s_{12})+\frac{1}{2}A_x(s,s_{23})
+\frac{1}{4}A_z(s,s_{23})+\frac{1}{4}A_w(s,s_{23})\right]\\
\\
A_w(s,s_{23})=\lambda\, b_w(s_{23})L_w(s_{23})+
K_w(s_{23})\left[\frac{1}{2}A_x(s,s_{12})
+\frac{1}{4}A_y(s,s_{12})+
\right.\\
\\
\hskip10ex \left.
+\frac{1}{4}A_v(s,s_{12})+\frac{1}{2}A_x(s,s_{13})
+\frac{1}{4}A_y(s,s_{13})+\frac{1}{4}A_v(s,s_{13})\right]\, .\\
\end{array}
\right.\eqno (A5)$$

\noindent
Here: for the $\Xi_{sb}^A$: $x=0^{qs}$, $y=0^{qb}$, $z=0^{sb}$,
$v=1^{qb}$, $w=1^{sb}$;
for the $\Xi_{cb}^A$: $x=0^{qc}$, $y=0^{qb}$, $z=0^{cb}$,
$v=1^{qb}$, $w=1^{cb}$;
for the $\Lambda_{scb}^A$: $x=0^{sc}$, $y=0^{sb}$, $z=0^{cb}$,
$v=1^{sb}$, $w=1^{cb}$;
for the $\Xi_{sb}^S$: $x=1^{qs}$, $y=1^{qb}$, $z=1^{sb}$,
$v=0^{qb}$, $w=0^{sb}$;
for the $\Xi_{cb}^S$: $x=1^{qc}$, $y=1^{qb}$, $z=1^{cb}$,
$v=0^{qb}$, $w=0^{cb}$;
for the $\Lambda_{scb}^S$: $x=1^{sc}$, $y=1^{sb}$, $z=1^{cb}$,
$v=0^{sb}$, $w=0^{cb}$.

\vskip2ex
{\bf Appendix B. Couple systems of approximate equations for the
${\bf S}$-wave bottom baryons.}
\vskip2ex
{\bf The $J^P=\frac{3}{2}^+$ multiplet.}
\vskip2ex

\noindent
1. Baryons $\Sigma_b$, $\Omega_{ssb}$, $\Omega_{ccb}$, $\Xi_{bb}$,
$\Omega_{sbb}$, $\Omega_{cbb}$.

$$\left\{
\begin{array}{l}
\alpha_{1^x}(s,s_0)=\lambda+2\, \alpha_{1^y}(s,s_0)
\, I_{1^x 1^y}(s,s_0)\, \frac{b_{1^y}(s_0)}{b_{1^x}(s_0)}\\
\\
\alpha_{1^y}(s,s_0)=\lambda+\alpha_{1^x}(s,s_0)
\, I_{1^y 1^x}(s,s_0)\, \frac{b_{1^x}(s_0)}{b_{1^y}(s_0)}
+\alpha_{1^y}(s,s_0)\, I_{1^y 1^y}(s,s_0)
\hskip2mm .\\
\end{array} \right.\eqno (A6)$$

\noindent
Here: for the $\Sigma_b$: $x=q$, $y=b$;
for the $\Omega_{ssb}$: $x=s$, $y=b$; for the $\Omega_{ccb}$: $x=c$, $y=b$;
for the $\Xi_{bb}$: $x=b$, $y=q$; for the $\Omega_{sbb}$: $x=b$, $y=s$;
for the $\Omega_{sbb}$: $x=b$, $y=c$.

\newpage
\noindent
2. Baryons $\Xi_{sb}$, $\Xi_{cb}$, $\Omega_{scb}$.

$$\left\{
\begin{array}{l}
\alpha_{1^x}(s,s_0)=\lambda+\alpha_{1^y}(s,s_0)
\, I_{1^x 1^y}(s,s_0)\, \frac{b_{1^y}(s_0)}{b_{1^x}(s_0)}
+\alpha_{1^z}(s,s_0)\, I_{1^x 1^z}(s,s_0)\,
\frac{b_{1^z}(s_0)}{b_{1^x}(s_0)}\\
\\
\alpha_{1^y}(s,s_0)=\lambda+\alpha_{1^x}(s,s_0)
\, I_{1^y 1^x}(s,s_0)\, \frac{b_{1^x}(s_0)}{b_{1^y}(s_0)}
+\alpha_{1^z}(s,s_0)\, I_{1^y 1^z}(s,s_0)\,
\frac{b_{1^z}(s_0)}{b_{1^y}(s_0)}\\
\\
\alpha_{1^z}(s,s_0)=\lambda+\alpha_{1^x}(s,s_0)
\, I_{1^z 1^x}(s,s_0)\, \frac{b_{1^x}(s_0)}{b_{1^z}(s_0)}
+\alpha_{1^y}(s,s_0)\, I_{1^z 1^y}(s,s_0)\,
\frac{b_{1^y}(s_0)}{b_{1^z}(s_0)}\, .\\
\end{array} \right.\eqno (A7)$$

\noindent
Here: for the $\Xi_{sb}$: $x=q$, $y=s$, $z=b$;
for the $\Xi_{cb}$: $x=q$, $y=c$, $z=b$;
for the $\Omega_{scb}$: $x=s$, $y=c$, $z=b$.

\vskip2ex

\noindent
3. Baryon $\Omega_{bbb}$.

$$
\begin{array}{l}
\alpha_{1^{bb}}(s,s_0)=\lambda+2\, \alpha_{1^{bb}}(s,s_0)
\, I_{1^{bb} 1^{bb}}(s,s_0)\, .\\
\end{array}
\eqno (A8)$$

\vskip2ex
{\bf The $J^P=\frac{1}{2}^+$ multiplet.}
\vskip2ex

\noindent
1. Baryons $\Sigma_b$, $\Lambda_b$, $\Omega_{ssb}$, $\Omega_{ccb}$,
$\Xi_{bb}$, $\Omega_{sbb}$, $\Omega_{cbb}$.

$$\left\{
\begin{array}{l}
\alpha_x(s,s_0)=\lambda+\frac{1}{2}\, \alpha_y(s,s_0)\, I_{xy}(s,s_0)
\, \frac{b_y(s_0)}{b_x(s_0)}
+\frac{3}{2}\, \alpha_z(s,s_0)\, I_{xz}(s,s_0)
\, \frac{b_z(s_0)}{b_x(s_0)}\\
\\
\alpha_y(s,s_0)=\lambda+\alpha_x(s,s_0)\, I_{yx}(s,s_0)
\, \frac{b_x(s_0)}{b_y(s_0)}
-\frac{1}{2}\, \alpha_y(s,s_0)\, I_{yy}(s,s_0)\\
\\
\hskip10ex
+\frac{3}{2}\, \alpha_z(s,s_0)\, I_{yz}(s,s_0)\, \frac{b_z(s_0)}{b_y(s_0)}\\
\\
\alpha_z(s,s_0)=\lambda+\alpha_x(s,s_0)\, I_{zx}(s,s_0)
\, \frac{b_x(s_0)}{b_z(s_0)}
+\frac{1}{2}\, \alpha_y(s,s_0)\, I_{zy}(s,s_0)
\, \frac{b_y(s_0)}{b_z(s_0)}\\
\\
\hskip10ex
+\frac{1}{2}\, \alpha_z(s,s_0)\, I_{zz}(s,s_0)
\, .\\
\end{array} \right.\eqno (A9)$$

\noindent
Here: for the $\Sigma_b$: $x=1^{qq}$, $y=1^{qb}$, $z=0^{qb}$;
for the $\Lambda_b$: $x=0^{qq}$, $y=0^{qb}$, $z=1^{qb}$;
for the $\Omega_{ssb}$: $x=1^{ss}$, $y=1^{sb}$, $z=0^{sb}$;
for the $\Omega_{ccb}$: $x=1^{cc}$, $y=1^{cb}$, $z=0^{cb}$;
for the $\Xi_{bb}$: $x=1^{bb}$, $y=1^{qb}$, $z=0^{qb}$;
for the $\Omega_{sbb}$: $x=1^{bb}$, $y=1^{sb}$, $z=0^{sb}$;
for the $\Omega_{cbb}$: $x=1^{bb}$, $y=1^{cb}$, $z=0^{cb}$.

\newpage

\noindent
2. Baryons $\Xi_{sb}^A$, $\Xi_{sb}^S$, $\Lambda_{scb}^A$, $\Lambda_{scb}^S$,
$\Xi_{cb}^A$, $\Xi_{cb}^S$.

$$\left\{
\begin{array}{l}
\alpha_x(s,s_0)=\lambda+\frac{1}{4}\, \alpha_y(s,s_0)\, I_{xy}(s,s_0)
\, \frac{b_y(s_0)}{b_x(s_0)}
+\frac{1}{4}\, \alpha_z(s,s_0)\, I_{xz}(s,s_0)
\, \frac{b_z(s_0)}{b_x(s_0)}\\
\\
\hskip10ex
+\frac{3}{4}\, \alpha_v(s,s_0)\, I_{xv}(s,s_0)
\, \frac{b_v(s_0)}{b_x(s_0)}
+\frac{3}{4}\, \alpha_w(s,s_0)\, I_{xw}(s,s_0)
\, \frac{b_w(s_0)}{b_x(s_0)}\\
\\
\alpha_y(s,s_0)=\lambda+\alpha_x(s,s_0)\, I_{yx}(s,s_0)
\, \frac{b_x(s_0)}{b_y(s_0)}
-\frac{1}{2}\, \alpha_z(s,s_0)\, I_{yz}(s,s_0)\, \frac{b_z(s_0)}{b_y(s_0)}\\
\\
\hskip10ex
+\frac{3}{2}\, \alpha_w(s,s_0)\, I_{yw}(s,s_0)\, \frac{b_w(s_0)}{b_y(s_0)}\\
\\
\alpha_z(s,s_0)=\lambda+\alpha_x(s,s_0)\, I_{zx}(s,s_0)
\, \frac{b_x(s_0)}{b_z(s_0)}
-\frac{1}{2}\, \alpha_y(s,s_0)\, I_{zy}(s,s_0)\, \frac{b_y(s_0)}{b_z(s_0)}\\
\\
\hskip10ex
+\frac{3}{2}\, \alpha_v(s,s_0)\, I_{zv}(s,s_0)\, \frac{b_v(s_0)}{b_z(s_0)}\\
\\
\alpha_v(s,s_0)=\lambda+\alpha_x(s,s_0)\, I_{vx}(s,s_0)
\, \frac{b_x(s_0)}{b_v(s_0)}
+\frac{1}{2}\, \alpha_z(s,s_0)\, I_{vz}(s,s_0)
\, \frac{b_z(s_0)}{b_v(s_0)}\\
\\
\hskip10ex
+\frac{1}{2}\, \alpha_w(s,s_0)\, I_{vw}(s,s_0)\, \frac{b_w(s_0)}{b_v(s_0)}\\
\\
\alpha_w(s,s_0)=\lambda+\alpha_x(s,s_0)\, I_{wx}(s,s_0)
\, \frac{b_x(s_0)}{b_w(s_0)}
+\frac{1}{2}\, \alpha_y(s,s_0)\, I_{wy}(s,s_0)
\, \frac{b_y(s_0)}{b_w(s_0)}\\
\\
\hskip10ex
+\frac{1}{2}\, \alpha_v(s,s_0)\, I_{wv}(s,s_0)\, \frac{b_v(s_0)}{b_w(s_0)}
\, .\\
\end{array} \right.\eqno (A10)$$

\noindent
Here: for the $\Xi_{sb}^A$: $x=0^{qs}$, $y=0^{qb}$, $z=0^{sb}$,
$v=1^{qb}$, $w=1^{sb}$;
for the $\Xi_{cb}^A$: $x=0^{qc}$, $y=0^{qb}$, $z=0^{cb}$,
$v=1^{qb}$, $w=1^{cb}$;
for the $\Lambda_{scb}^A$: $x=0^{sc}$, $y=0^{sb}$, $z=0^{cb}$,
$v=1^{sb}$, $w=1^{cb}$;
for the $\Xi_{sb}^S$: $x=1^{qs}$, $y=1^{qb}$, $z=1^{sb}$,
$v=0^{qb}$, $w=0^{sb}$;
for the $\Xi_{cb}^S$: $x=1^{qc}$, $y=1^{qb}$, $z=1^{cb}$,
$v=0^{qb}$, $w=0^{cb}$;
for the $\Lambda_{scb}^S$: $x=1^{sc}$, $y=1^{sb}$, $z=1^{cb}$,
$v=0^{sb}$, $w=0^{cb}$.

\newpage

\noindent
Table I. Bottom baryon masses of multiplet $\frac{1}{2}^+$.

\vskip1.5ex
\noindent
Parameters of model: quark masses $m_{u,d}=495\, MeV$, $m_s=770\, MeV$,
$m_c=1655\, MeV$, $m_b=4840\, MeV$; cutoff parameters: $\lambda_q=10.7$
($q=u, d, s$), $\lambda_c=6.5$, $\lambda_b=5.4$; gluon coupling constants:
$g_0=0.70$, $g_1=0.55$ with $J^p=0^+$ and $1^+$, $g_c=0.857$, $g_b=1.03$.

\vskip1.5ex

\noindent
\begin{tabular}{|c|c|c|}
\hline
Baryon & Mass ($GeV$) & Mass ($GeV$) (exp.)\\
\hline
$\Sigma_b$ & $5.808$ & $5.808$ \\
\hline
$\Lambda_b$ & $5.624$ & $5.624$ \\
\hline
$\Xi_{sb}^A$ & $5.761$ & $5.793$ \\
\hline
$\Xi_{sb}^S$ & $6.007$ & -- \\
\hline
$\Omega_{ssb}$ & $6.120$ & -- \\
\hline
$\Xi_{cb}^A$ & $6.789$ & -- \\
\hline
$\Xi_{cb}^S$ & $6.818$ & -- \\
\hline
$\Lambda_{scb}^A$ & $6.798$ & -- \\
\hline
$\Lambda_{scb}^S$ & $6.836$ & -- \\
\hline
$\Omega_{ccb}$ & $7.943$ & -- \\
\hline
$\Xi_{bb}$ & $10.045$ & -- \\
\hline
$\Omega_{sbb}$ & $9.999$ & -- \\
\hline
$\Omega_{cbb}$ & $11.089$ & -- \\
\hline
\end{tabular}

\vskip8ex

\noindent
Table II. Bottom baryon masses of multiplet $\frac{3}{2}^+$.

\vskip1.5ex

\noindent
\begin{tabular}{|c|c|c|}
\hline
Baryon & Mass ($GeV$) & Mass ($GeV$) (exp.)\\
\hline
$\Sigma_b$ & $5.829$ & $5.829$ \\
\hline
$\Xi_{sb}$ & $6.066$ & -- \\
\hline
$\Omega_{ssb}$ & $6.220$ & -- \\
\hline
$\Xi_{cb}$ & $6.863$ & -- \\
\hline
$\Omega_{scb}$ & $6.914$ & -- \\
\hline
$\Omega_{ccb}$ & $7.973$ & -- \\
\hline
$\Xi_{bb}$ & $10.104$ & -- \\
\hline
$\Omega_{sbb}$ & $10.126$ & -- \\
\hline
$\Omega_{cbb}$ & $11.123$ & -- \\
\hline
$\Omega_{bbb}$ & $14.197$ & -- \\
\hline
\end{tabular}

\vskip10ex
\noindent
Table III. Coefficients of Ghew-Mandelstam functions.

\vskip1.5ex

\begin{tabular}{|c|c|c|c|}
\hline
 &$\alpha_J$&$\beta_J$&$\delta_J$\\
\hline
 & & & \\
$1^+$&$\frac{1}{3}$&$\frac{4m_i m_k}{3(m_i+m_k)^2}-\frac{1}{6}$
&$-\frac{1}{6}(m_i-m_k)^2$\\
 & & & \\
$0^+$&$\frac{1}{2}$&$-\frac{1}{2}\frac{(m_i-m_k)^2}{(m_i+m_k)^2}$&0\\
 & & & \\
\hline
\end{tabular}

\newpage
{\bf \Large References.}
\vskip5ex

\noindent
1. N. Isgur and M.B. Wise, Phys. Lett. B232, 113 (1989).

\noindent
2. H. Georgi, Phys. Lett. B240, 447 (1990).

\noindent
3. A.F. Falk, H. Georgi, B. Grinstein and M.B. Wise,
Nucl. Phys. B343, 1 (1990).

\noindent
4. T. Aaltonen, et al., CDF Collaboration, arXiv: 0706.3868 [hep-ex].

\noindent
5. V. Abazov et al., D0 Collaboration, Phys. Rev. Lett. 99, 052001 (2007).

\noindent
6. T. Aaltonen et al., CDF Collaboration, Phys. Rev. Lett. 99, 052002
(2007).

\noindent
7. S. Capstick and N. Isgur, Phys. Rev. D34, 2809 (1986).

\noindent
8. R. Roncaglia, D.B. Lichtenberg and E. Predazzi, Phys.
Rev. D52, 1722 (1995).

\noindent
9. E.E. Jenkins, Phys. Rev. D54, 4515 (1996).

\noindent
10. N. Mathur, R. Lewis and R.M. Woloshyn, Phys. Rev. D66, 014502 (2002).

\noindent
11. D. Ebert, R.N. Faustov and V.O. Galkin, Phys. Rev. D72, 034026 (2005).

\noindent
12. M. Karliner and H.J. Lipkin, arXiv: hep-ph/0611306.

\noindent
13. M. Karliner, B. Keren-Zur, H.J. Lipkin and J.L. Rosner,
arXiv: 0706.2163

[hep-ph].

\noindent
14. M. Karliner, B. Keren-Zur, H.J. Lipkin and J.L. Rosner,
arXiv:0708.4027 [hep-ph].

\noindent
15. J. L. Rosner, Phys. Rev. D75, 013009 (2007).

\noindent
16. M. Karliner and H.J. Lipkin, Phys. Lett. B575, 249 (2003).

\noindent
17. C. Chen, X.L. Chen, X. Liu, W.Z. Deng and S.L. Zhu,
Phys. Rev. D75, 094017

(2007).

\noindent
18. E. Bagan, M. Chabab, H.G. Dosch and S. Narison, Phys.
Lett. B278, 367 (1992).

\noindent
19. D.W. Wang, M.Q. Huang and C.Z. Li, Phys. Rev. D65, 094036 (2002).

\noindent
20. S.L. Zhu, Phys. Rev. D61, 114019 (2000).

\noindent
21. F.O. Duraes and M. Nielsen, arXiv:0708.3030 [hep-ph].

\noindent
22. X. Liu, H.X. Chen, Y.R. Lui, A. Hosaka and S.L. Zhu,
arXiv:0710.0123 [hep-ph].

\noindent
23. S.M. Gerasyuta, Z. Phys. C60, 683 (1993).

\noindent
24. S.M. Gerasyuta, N. Cim. A106, 37 (1993).

\noindent
25. S.M. Gerasyuta and D.V. Ivanov, N. Cim. A112, 261 (1999).

\noindent
26. A.De Rujula, H.Georgi and S.L.Glashow, Phys. Rev. D12, 147 (1975).

\noindent
27. V.V. Anisovich, S.M. Gerasyuta, and A.V. Sarantsev,
Int. J. Mod. Phys. A6, 625

(1991).

\noindent
28. G. Chew, S. Mandelstam, Phys. Rev. 119, 467 (1960).

\noindent
29. V.V. Anisovich and A.A. Anselm, Usp. Fiz. Nauk 88, 287 (1966).

\noindent
30. W.M. Yao et al. (Particle Data Group), J. Phys. G33, 1 (2006) and

2007 partial update for edition 2008.

\noindent
31. M. Karliner, B. Keren-Zur, H.J. Lipkin, and J.L. Rosner,
arXiv:0804.1575 [hep-ph].

\noindent
32. I.M. Narodetski, M.A. Trusov and A.I. Veselov, arXiv:0801.1980 [hep-ph].

\noindent
33. W. Roberts and M. Pervin, arXiv:0711.2492 [nucl-th].

\noindent
34. E.E. Jenkins, Phys. Rev. D77, 034012 (2008).

\noindent
35. S.M. Gerasyuta and E.E. Matskevich, Phys. Rev. D76, 116004 (2007).

\noindent
36. S.M. Gerasyuta and E.E. Matskevich, Yad. Fiz. 70, 1995 (2007)
[Phys. At. Nucl.

70, 1946 (2007)].

\noindent
37.  S. Capstick et al., arXiv:0711.1982 [hep-ph].

\end{document}